# Correlation between $T_c$ and the Pseudogap Observed in the Optical Spectra of High $T_c$ Superconducting Cuprates

Setsuko Tajima[1]*, Yuhta Itoh[1], Katsuya Mizutamari[1], Shigeki Miyasaka[1], Masamichi Nakajima[1], Nae Sasaki[2], Shunpei Yamaguchi[2], Kei-ichi Harada[2], and Takao Watanabe[2]

[1]*Department of Physics, Osaka University, Toyonaka, Osaka 560-0043, Japan*
[2]*Graduate School of Science and Technology, Hirosaki University, Aomori 036-8561, Japan*

We studied the temperature dependences of the optical spectra for optimally and underdoped $Bi_2Sr_2Ca_2Cu_3O_{10+z}$ single crystals. Similarly to the other cuprates' cases, a gap-like conductivity suppression was observed with reducing the temperature from above $T_c$, creating a peak in the conductivity spectrum. The conductivity peak energy was insensitive to the doping level, namely $T_c$, which suggests that this gap is not a superconducting gap but is related to the pseudogap. Comparing the data of various mono-, double-, and triple-layer cuprates, we found a clear correlation between the optimal $T_c$ of each material and the pseudogap-related conductivity peak energy.

Despite significant efforts since 1986, the superconductivity mechanism of high-temperature superconducting cuprates remains a subject of debate. The electronic phase diagram has become rich due to various coexisting orders discovered recently[1,2], however, the pseudogap has been a central issue in understanding the electronic state of this material system[3]. It is apparent that the pseudogap is competing with superconductivity because it reduces the effective Fermi surface and decreases the free carrier density[4]. As the carrier doping level is reduced, the pseudogap temperature increases while the superconducting temperature $T_c$ decreases. If the pseudogap competes with superconductivity, it is unlikely to act as a glue for superconducting pairing. Despite this, we cannot conclude that the pseudogap is irrelevant to superconductivity because both superconductivity and the pseudogap could originate from a strong electron correlation.

Determining the gap opening temperature is generally easier than determining the gap energy because this can be done by measuring the temperature dependencies of physical quantities. Unlike a superconducting gap, the pseudogap does not show a sharp structure in excitation spectra but rather gradually suppresses low energy excitation. Due to these issues, there has been a limited systematic research on the pseudogap energy, including its dependence on doping, material, and the number of $CuO_2$-layer, except for the angle-resolved photoemission study[5]. The relationship between the pseudogap size and the superconductivity transition temperature ($T_c$) is not yet clear, although it is crucially important to discuss the role of the pseudogap in high $T_c$ superconductivity.

Optical spectroscopy is a powerful experimental tool used to determine the gap value. In most cuprates, a gap-like structure was observed in optical conductivity spectra at low temperatures.[6-8] In most cases, the suppression of conductivity begins at a temperature significantly higher than $T_c$, as was already reported at the very beginning of high $T_c$ superconductivity research.[9] Therefore, it is natural to consider that this structure represents the pseudogap rather than the superconducting gap. However, there are some other interpretations that this structure is related to superconductivity, for example, the superconducting gap itself[10] or the energy of a boson coupled to electrons.[11,12]

While most of the typical high $T_c$ cuprates such as $Bi_2Sr_2CaCu_2O_{8+z}$ (Bi2212) and $YBa_2Cu_3O_{7-\delta}$ (Y123) with $T_c$ ~90 K commonly showed the optical gap-like structure at around 1000 cm$^{-1}$,[9,13-15] a larger energy scale (~1200 cm$^{-1}$) was reported in the triple layer $HgBa_2Ca_2Cu_3O_{8+z}$ (Hg1223) with $T_c$ =121 K.[16] To identify the source of this high energy value of the gap, it is necessary to examine the optical spectra for other triple-layers cuprates with $T_c$ higher than 90 K. In this work, we investigate the in-plane optical spectra of another triple layer compound $Bi_2Sr_2Ca_2Cu_3O_{10+z}$ (Bi2223), and discuss the dependence of the gap size on doping level, material, and the number of $CuO_2$-layer.

The single crystals of Bi2223 were grown by a traveling solvent floating zone method as described elsewhere.[17, 18] In the present study, we used two pieces of crystals taken from the same crystal rod. The sample sizes are 1 x 1.5 x 0.5 mm$^3$ and 1.5 x 2 x 0.5 mm$^3$, respectively. The first piece is optimally doped with $T_c$ =109 K (OP109), while the second is the underdoped one with $T_c$ =88K (UD88). Here the $T_c$ values were determined from the temperature dependence of magnetic susceptibility. The carrier doping level ($p$) was controlled by changing oxygen content through annealing.[19] Here, $p$ is defined by the average Cu valence 2+$p$. It should be noted that in the triple layer cuprates such as Bi2223, the carrier doping level is different between the outer and inner $CuO_2$-planes.[20] In the case of OP109, the average doping level is $p_{avr}$ = 0.16, but the NMR experiment revealed that the outer plane is slightly overdoped ($p$=0.203) and the inner plane is underdoped ($p$ = 0.127).[21] Similarly, for UD88 with $p_{avr}$ = 0.11, the doping levels of the outer and the inner planes were estimated to be $p$ = 0.155 and 0.079, respectively.[22] The Raman scattering spectra revealed two peaks corresponding to the two superconducting gaps for the differently-doped planes.[22] However, the optical spectra have never shown the coexisting two gaps, because the larger energy gap must dominate the spectral feature.

The samples were cleaved just before being inserted into the cryostat. We measured optical reflectivity spectra from 100 to 20000 cm$^{-1}$ using a Fourier transform spectrometer equipped with a microscope and a cryostat. Between 500 and 20000 cm$^{-1}$, the microscope with a cooling stage was used. Here, gold was evaporated on half of the sample surface which was used as a reference to estimate reflectivity. For far-infrared measurements between 100 and 2000 cm$^{-1}$, we used a standard setup with a He-flow cryostat instead of the microscope system. We estimated the absolute value of reflectivity, based on the value determined by microscope measurements. For the Kramers-Kronig transformation, we also used the high energy data above 20000 cm$^{-1}$ which was measured at the synchrotron facility, UVSOR in Okazaki, Japan. The low energy extrapolation was made using the Hagen-Rubens equation.

Figure 1 shows the in-plane polarized reflectivity spectra of two Bi2223 samples at various temperatures. There is no significant difference between the spectra of the two samples. In both cases, the reflectivity increases as the frequency decreases below 10000 cm$^{-1}$, forming a Drude-like reflectivity edge structure. With decreasing temperature, the low-frequency reflectivity monotonically increases, which is a typical metallic behavior. Significant reflectivity changes with $T$ are visible below ~1000 cm$^{-1}$, while the $T$-dependence is weak above ~1000 cm$^{-1}$. The low-$T$ spectra reveal a kink structure, corresponding to the gap opening, as seen in the figures' insets.

The optical conductivity spectra were calculated from the reflectivity spectra via the Kramers-Kronig transformation, as shown in Fig.2. Both the spectra for the optimally and under-doped samples are similar. Namely, the conductivity increases towards zero frequency at high temperatures, while it is suppressed below 1200 cm$^{-1}$ at low temperatures. The spectrum for OP109 agrees with the spectra reported by Carbone $et\ al.$[10] The conductivity suppression is considered as an effect of the gap opening. When we analyze the spectra by using the extended Drude model, we obtain the frequency-dependent scattering rate $1/\tau(\omega)$. The gap opening behavior in $\sigma_1(\omega)$ can be seen as the downward bending of $1/\tau(\omega)$ from the $\omega$-linear behavior. The frequency of this kink in $1/\tau(\omega)$ is also a measure of the gap energy.

The gap opening temperature refers to the temperature at which the spectral shape qualitatively changes from a monotonically increasing spectrum to one showing a peak around 1200 cm$^{-1}$. We found that the threshold temperature is approximately 130 K for OP109 and 150 K for UD88 in Fig.2. The dc resistivity $\rho(T)$ of each sample also exhibits a downward bending around this threshold temperature, respectively.[19]

The conductivity suppression at low frequencies can be seen in the change of spectral weight (SW). Figure 3 demonstrates the $T$-dependence of the low $\omega$-SW normalized at the room temperature. The SW between $\omega_1$ and $\omega_2$ is defined by the following integral,

$$\text{SW}=\int_{\omega_1}^{\omega_2}\sigma_1(\omega)d\omega\ . \qquad (1)$$

As shown in Fig.3, for OP109, both the low $\omega$-SW ($\omega_2$=2000 cm$^{-1}$) and the high $\omega$-SW ($\omega_2$=9500 cm$^{-1}$) begin to decrease around 130K, corresponding to the gap-opening temperature. The changing rate is larger in the low $\omega$-SW than the high $\omega$-SW, implying that the SW suppression mainly occurs in the low $\omega$ region. In the case of UD88, the $T$-dependence of the SW indicates that the gap-opening temperature is around 150K.

It is important to note that both temperatures are higher than the $T_c$ value of each sample, respectively. This implies that the observed conductivity peaks are not due to a superconducting gap but rather due to something else such as the pseudogap.

Another important fact in Fig.2 is that the conductivity peak frequency is very similar in the two samples despite their different doping levels and different $T_c$ values. It means that the observed energy scale is insensitive to the doping level. Although one may speculate that it may be due to the coexistence of the differently doped inner- and outer $CuO_2$ layers in Bi2223, the doping insensitive behavior of the conductivity peak has been commonly observed in the optical spectra of various high $T_c$ cuprates, as noted by many researchers.[6,9] This further supports the idea that the observed gap-like feature is not actually a superconducting gap.

The conductivity peak energy is less affected by doping, but it seems to depend on the number (n) of the $CuO_2$-layers in a formula unit. Figure 4 compares the optical conductivity spectra of three Bi-based cuprates: $Bi_2Sr_2CuO_{6+z}$ (Bi2201, n = 1),[23] $Bi_2Sr_2CaCu_2O_{8+z}$ (Bi2212, n = 2),[15] and $Bi_2Sr_2Ca_2Cu_3O_{10+z}$ (Bi2223, n = 3). All three spectra display a broad peak but at different energies: around 650 cm$^{-1}$ (~81 meV) for Bi2201, 1000 cm$^{-1}$ (~124 meV) for Bi2212, and 1200 cm$^{-1}$ (~149 meV) for Bi2223. The pseudogap energy systematically increases with the number of the $CuO_2$-layer. This n-dependence is not accidental but rather common in the cuprates. For example, the typical monolayer cuprate, $(La, Sr)_2CuO_4$ (La214), also exhibits a peak in the conductivity spectrum around 650 cm$^{-1}$ at low temperatures.[24] As to the double-layer compounds, Y123[9,13,14] and $YBa_2Cu_4O_8$ (Y124)[6] show a peak in $\sigma_1(\omega)$ at around 1000 cm$^{-1}$, similar to Bi2212. While there are not many reports for triple-layer cuprates, a study of Hg1223 crystals with two doping levels showed that a conductivity peak was commonly observed around 1200 cm$^{-1}$ in both samples.[16]

Considering the conductivity peak energy $\Delta_g$ as a measure of the pseudogap energy, we plot $\Delta_g$ and $T_c^{max}$ ($T_c$ at the optimal doping) for various cuprates in Fig.5. One can see the clear correlation between $\Delta_g$ and $T_c^{max}$. Here we used $T_c^{max}$ rather than the number of the $CuO_2$-layers (n) as a key parameter determining $\Delta_g$. Plotting as a function of $T_c^{max}$ seems to be equivalent to that as a function of n because it is well known that $T_c^{max}$ increases with increasing n. However, there are important data points for $HgBa_2CuO_{4+z}$ (Hg1201)[12] and $TlBa_2CuO_{4+z}$ (Tl1201)[6] which are monolayer compounds but have a high $T_c$ of 93 K and 90 K, respectively. The pseudogap of these compounds is about 1000 cm$^{-1}$ which is close to the value of $\Delta_g$ for double-layer compounds. This means that $\Delta_g$ is dependent on $T_c$ but not on n.

A similar correlation between $T_c$ and the optical gap feature was pointed out in ref.12, although the authors interpreted this structure as being related to a peak in the electron-boson spectral density function $I^2\chi(\Omega)$. One of the clear differences from our observation is that the peak energy in $I^2\chi(\Omega)$ changes with carrier doping.

In the angle-resolved photoemission (ARPES) experiments, the pseudogap energy was defined as the gap energy ($\Delta^*$) extrapolated to the ($\pi$, 0) direction, which was distinct from the nodal direction gap.[5, 25] A weak correlation between $\Delta^*$ and $T_c$ was also found from the comparison of the data for single-, double- and triple-layer Bi-based cuprates. ($\Delta^*$ ~60 meV for n = 1, ~75 meV for n = 2, and ~80 meV for n = 3).[5] Since $\Delta^*$ is the maximum gap energy at the anti-nodal direction, it is reasonable that $2\Delta^*$ is larger than $\Delta_g$ determined by the optical conductivity peak that reflects the gap averaged over the entire Fermi surface.

It should be noted that the doping-insensitive behavior of $\Delta_g$ indicates that it is different from the pseudogap observed in ARPES and tunneling spectroscopy. The pseudogap studies by tunneling technique, though they were limited to the Bi-based cuprates, demonstrated that the pseudogap energy decreases with carrier doping.[26-28] This technique mainly monitors the gap in the anti-nodal direction, while the in-plane optical spectra are dominated by the coherent charge dynamics of the carriers in the nodal region. In optical spectroscopy, the anti-nodal gap can be detected with the c-axis polarized light. Actually, the spectral weight transfer due to the pseudogap was observed to extend to very high energies such as 5000 cm$^{-1}$ in the c-axis spectra of Y123.[29] Then, what is the gap-like structure in the in-plane optical spectra? It manifests itself as a result of the radical Drude narrowing below the pseudogap temperature. It may originate from some excitation or the change in the $\omega$-dependent scattering rate. If it is related to the so-called nematic order or the charge density wave,[1,2] the effect of such an order on the in-plane and the inter-plane charge dynamics could be different.

Finally, we discuss the meaning of the result in Fig.5. Why the $T_c$ values of La214, Y123, and Hg1223 are so different despite sharing the same $CuO_2$ physics is still an open question. The idea that the $T_c$ variation is caused by disorders or distortion of the $CuO_2$-planes[30] is not successful in explaining the experimental facts. For example, judging from the optical spectra, we can conclude that the disorder effect in Y123 is weak because the optical spectra of optimally doped Y123 show little low-$\omega$ residual conductivity that is often observed in the impurity-doped cuprates.[14] Another idea is that competing orders suppress superconductivity and thus cause the decrease in $T_c$, but this is not appropriate, because a larger competing energy in a lower $T_c$ compound should give a larger pseudogap, which is opposite to the present observation. The theoretical model that the coexistence of the highly doped outer $CuO_2$-planes and the pseudogapped inner plane enhances the $T_c$ in the triple layer cuprates[31] seems interesting. However, it cannot explain the variation in the pseudogap as well as the correlation of $T_c^{max}$ and $\Delta_g$. One may remember that in the case of Fe-based superconductors, the local crystal structure such as the bond distance and/or the bond angle is a crucial parameter to determine the $T_c$ values.[32, 33] However, this is a multi-band system, and thus even a small change in local structure parameters can affect the contribution of each band to the Fermi surface, which could alter the crucial interaction for superconductivity.

Our findings of the correlation between $T_c^{max}$ and $\Delta_g$ in the present study is an experimental fact independent of any model. This indicates that the pseudogap is tightly linked to the superconductivity mechanism in the cuprates. In other words, the electronic interaction that determines the pseudogap-related energy $\Delta_g$ is also responsible for determining the $T_c$ value, although the pseudogap inhibits electron itinerancy. The microscopic theory to clarify the essential parameters that determine $\Delta_g$ is highly desired for elucidation of the high $T_c$ superconductivity mechanism.

In summary, we have measured the in-plane polarized optical reflectivity spectra of optimally and under-doped Bi2223 crystals at various temperatures. A gap-like conductivity suppression was observed at low temperatures below 1200 cm$^{-1}$ (~149 meV) in both samples. This gap energy was found to be doping-insensitive and the gap opening temperature was higher than $T_c$, which strongly suggests that the observed gap is not a superconducting gap but related to the pseudogap. Comparing the data from various mono-, double-, and triple-layer cuprates, we found a strong correlation between the gap energy and the optimal $T_c$. This correlation should be seriously taken into account in the theoretical models for high $T_c$ superconductivity of the cuprates.

**Acknowledgment**

This work was supported by KAKENHI (Grant Number 18K03513 and 25400349). The measurement of high energy optical spectra was supported by the Use-of-UVSOR Synchrotron Facility Program (Proposals No. 30-872) of the Institute for Molecular Science, Okazaki, Japan.

*E-mail: tajimasetsu@gmail.com

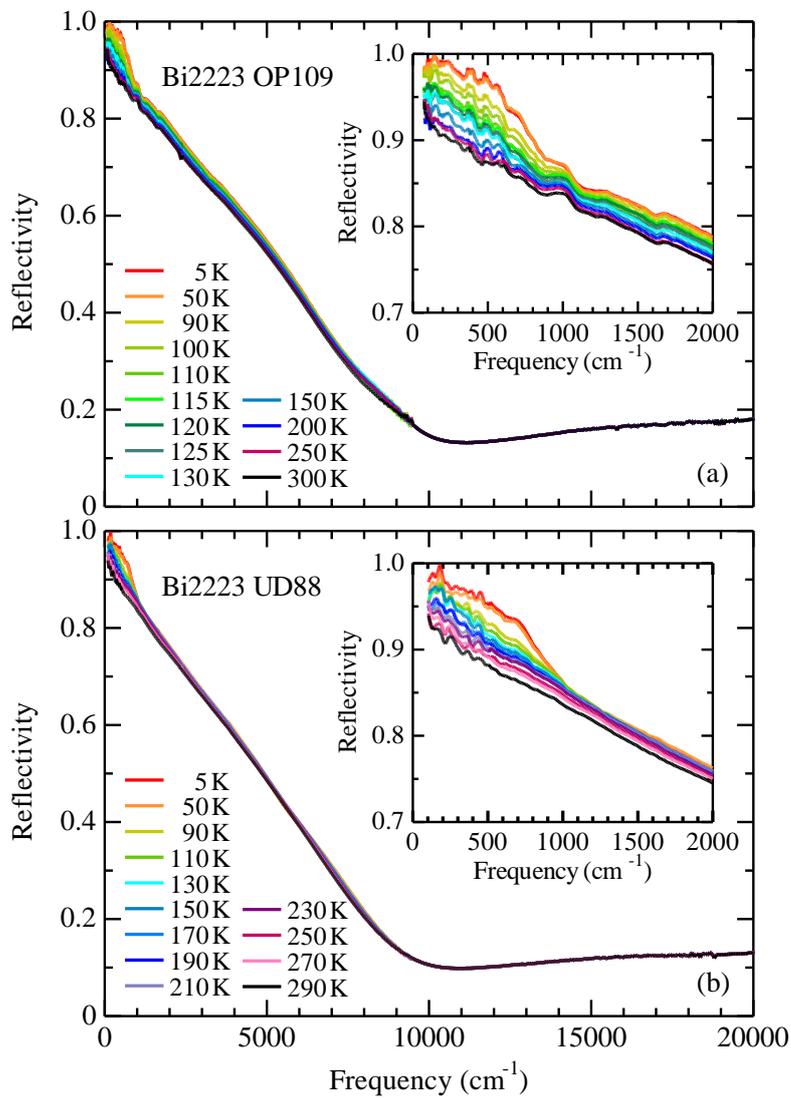

Fig.1. (Color online) Reflectivity spectra of $Bi_2Sr_2Ca_2Cu_3O_{10+z}$ (Bi2223) crystals with E//ab at various temperatures. (a) Optimally doped crystal with $T_c$=109K. (b) Under-doped crystal with $T_c$= 88K. The insets are the expanded scale spectra below 2000 cm$^{-1}$.

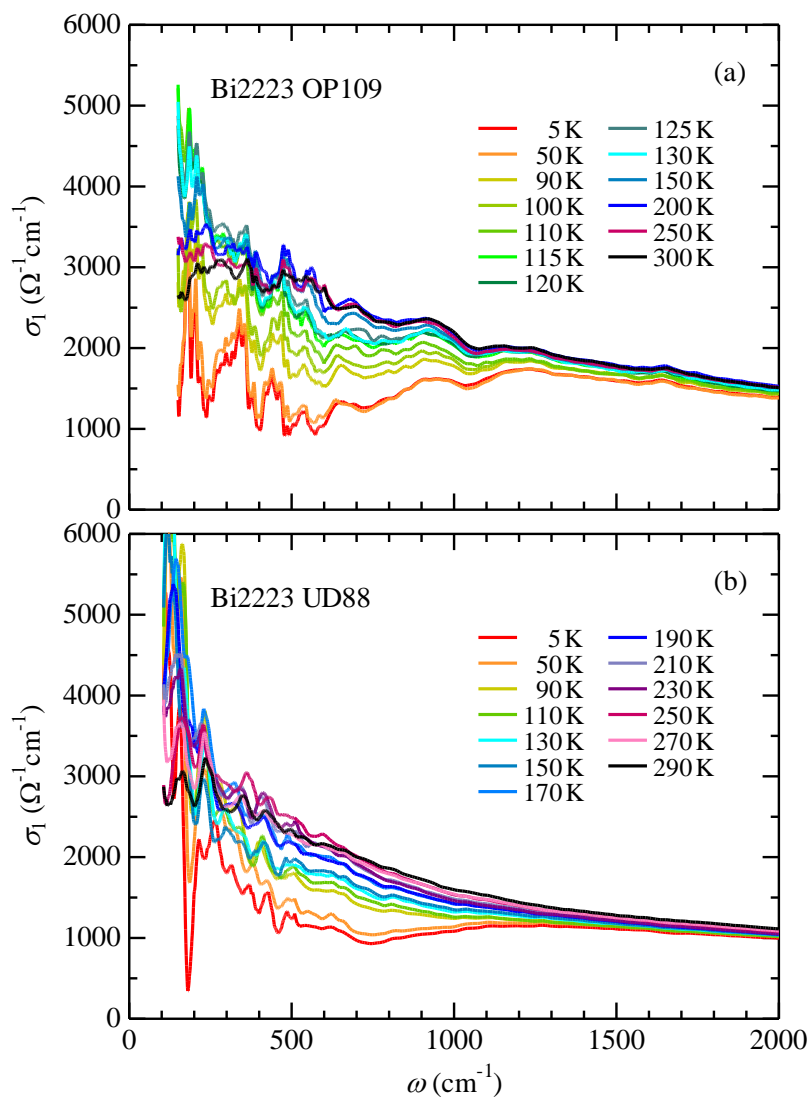

Fig.2. (Color online) In-plane conductivity spectra of Bi2223 crystals which were calculated from the reflectivity spectra in Fig.1

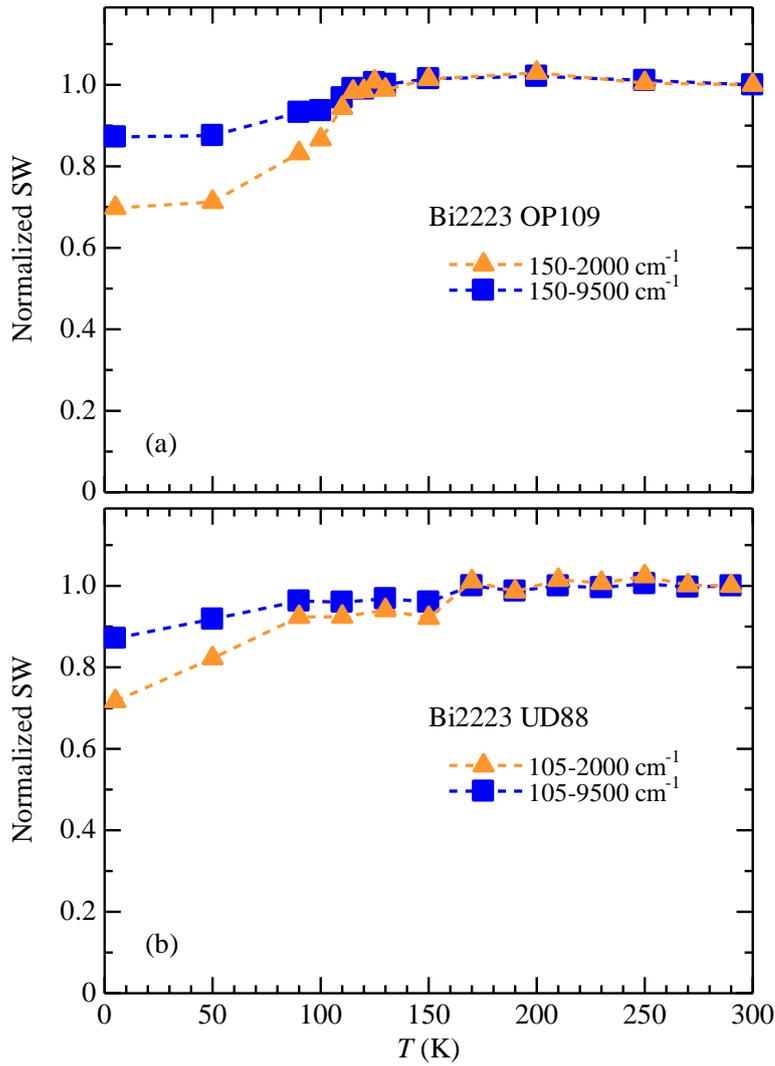

Fig. 3. (Color online) Temperature dependences of the normalized spectral weight (SW) in the conductivity spectra of optimally (a) and underdoped Bi2223 (b). The data are normalized at the room temperature. The triangles are the SWs between 150 (or 105) and 2000 cm$^{-1}$, and the squares are those between 150 (or 105) and 9500 cm$^{-1}$.

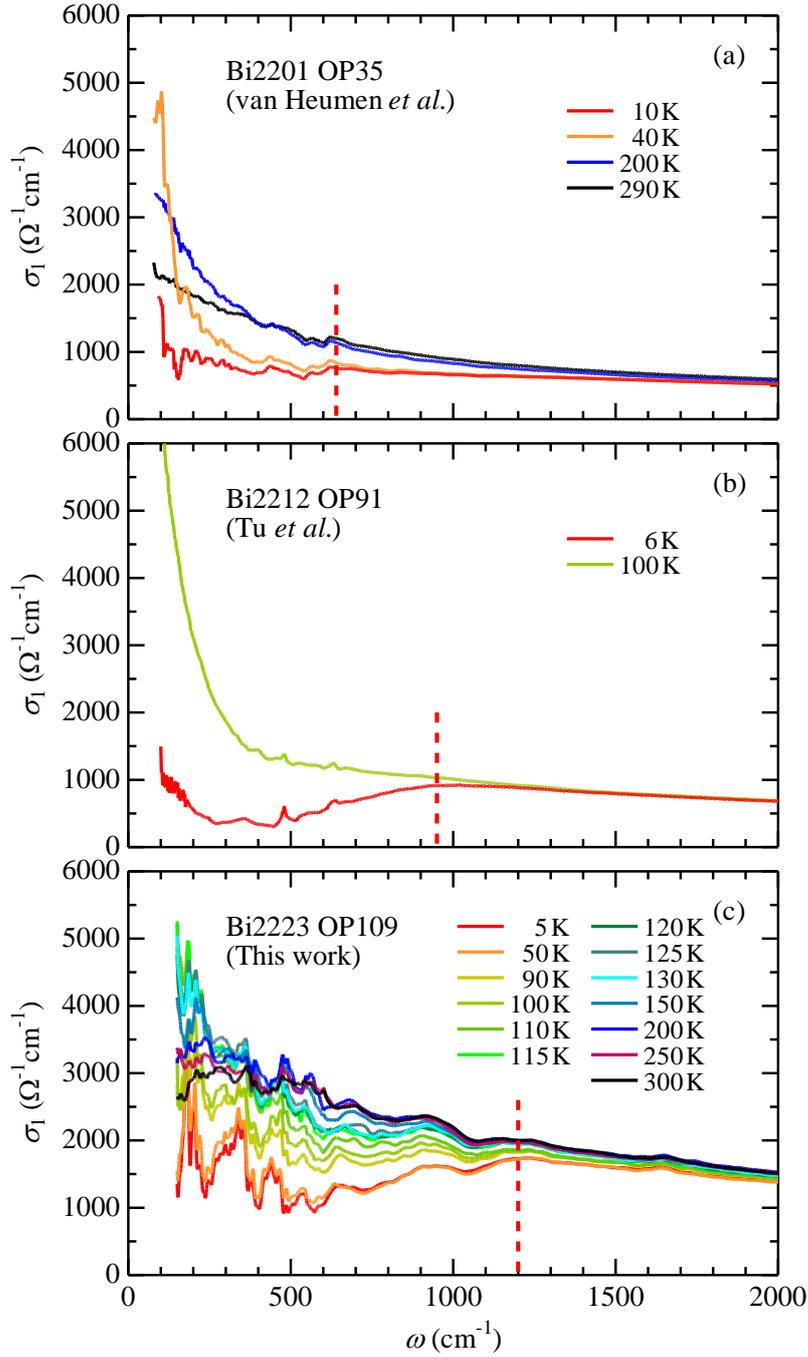

Fig.4. (Color online) Comparison of the in-plane conductivity spectra of single layer $Bi_2Sr_2CuO_{6+z}$ (Bi2201) with $T_c$=35K (a), double layer $Bi_2Sr_2CaCu_2O_{8+z}$ (Bi2212) with $T_c$=91K (b) and the present results of Bi2223 (c). The spectra of Bi2201 are taken from ref.23, and those of Bi2212 are from ref.15. The red dashed bars indicate the threshold frequencies at which the conductivity suppression starts.

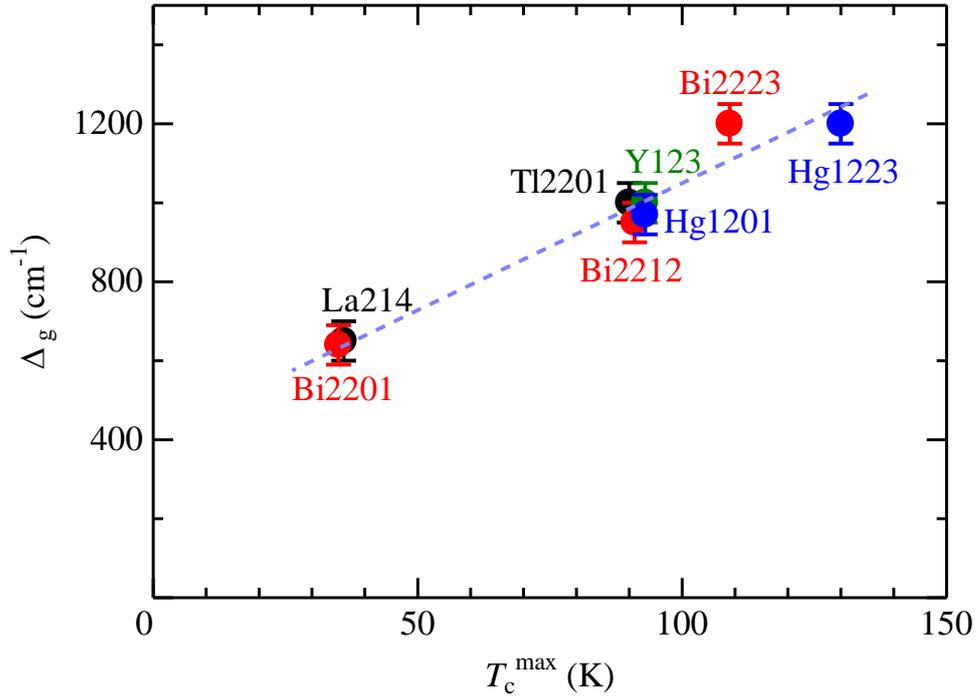

Fig.5. (Color online) Comparison of the gap energy $\Delta_g$ and $T_c^{max}$ for various copper oxide superconductors. $\Delta_g$ was defined by the threshold frequency for conductivity suppression as indicated for Bi2201, Bi2212, and Bi2223 in Fig.4. $T_c^{max}$ is the maximum $T_c$ for each compound at the optimum doping. The data of $(La,Sr)_2CuO_4$ (La214) is from ref. 24, $Tl_2Ba_2CuO_{6+z}$ (Tl2201) from ref. 6, $YBa_2Cu_3O_{7-\delta}$ (Y123) from ref.14, $HgBa_2CuO_{4+z}$ (Hg1201) from ref.12, $HgBa_2Ca_2Cu_3O_{8+z}$ (Hg1223) from ref. 16. The dashed line is a guide for eye.